# Growth of Transition Metal Sulfides by Sulfuric Vapor Transport and Liquid Sulfur: Synthesis and Properties

D. A. Chareev[1,2,3], D. Phyual[4], D. Karmakar[5], A. Nekrasov[1], F. O. L. Johansson[6,7,8], T. Sarkar[9], H. Rensmo[5], Olle Eriksson[5,10], Anna Delin[4,11], A.N. Vasiliev[12,2,3], Mahmoud Abdel-Hafiez[5]

[1] Institute of Experimental Mineralogy (IEM RAS), 142432 Chernogolovka, Moscow Region, Russia
[2] National University of Science and Technology "MISiS", Moscow 119049, Russia
[3] Ural Federal University, Ekaterinburg 620002, Russia
[4] Department of Applied Physics, KTH Royal Institute of Technology, SE-106 91 Stockholm, Sweden
[5] Department of Physics and Astronomy, Uppsala University, Box 516, SE-75120 Uppsala, Sweden
[6] Institute for Methods and Instrumentation in Synchrotron Radiation Research, PS-ISRR, Helmholtz-Zentrum Berlin für Materialien und Energie, Albert-Einstein-Strasse 15, 12489 Berlin, Germany
[7] Institut für Physik und Astronomie, Universität Potsdam, Karl-Liebknecht-Strasse 24, 14476 Potsdam, Germany
[8] Division of Applied Physical Chemistry, Department of Chemistry, KTH – Royal Institute of Technology, SE-100 44 Stockholm, Sweden
[9] Department of Materials Science and Engineering, Box 35, Uppsala University, SE-751 03, Sweden
[10] School of Science and Technology, Örebro University, SE-701 82 Örebro, Sweden
[11] Swedish e-Science Research Center, KTH Royal Institute of Technology, SE-10044 Stockholm, Sweden
[12] Lomonosov Moscow State University, Moscow 119991, Russia

**ABSTRACT**: Transition metals dichalcogenides (TMDs) are an emergent class of low-dimensional materials with growing applications in the field of nanoelectronics. However, efficient methods for synthesizing large mono-crystals of these systems are still lacking. Here, we describe an efficient synthetic route for a large number of TMDs that were obtained in quartz ampoules by sulfuric vapor transport and liquid sulfur. Crystals of metal sulfides MgS, PdS, $PtS_2$, $ReS_2$, $NbS_2$, $TaS_2$, $TaS_3$, $MoS_2$, $WS_2$, $FeS_2$, $CoS_2$, $NiS_2$, $Cr_2S_3$, $V_{1+\delta}S_2$, $In_2S_3$, $Bi_2S_3$, $TiS_2$, $ZrS_3$, $HfS_3$, and pure Au were obtained in quartz ampoules by chemical vapor transport technique with sulfur vapors as the transport agent. Unlike the sublimation technique, the metal enters the gas phase in the form of molecules, hence containing greater amount of sulfur than the growing crystal. We have investigated the physical properties for a selection of these crystals and compared them to state-of-the-art findings reported in the literature. The acquired x-ray photoemission spectroscopy features demonstrate the overall high quality of single crystals grown in this work as exemplified by $ReS_2$ and $CoS_2$. This new approach to synthesize high-quality transition metal dichalcogenides single crystals can alleviate many material quality concerns and is suitable for emerging electronic devices.

## Introduction

Two-dimensional (2D) transition metals dichalcogenides are an emergent class of materials with growing applications in the field of nanoelectronics [1]. Some examples of their use are in heterostructures and monolayers based on $MoS_2$ and $WS_2$ as transistors, $MoS_2$ and $MoTe_2$ as phototransistors, $WS_2$, $SnS_2$, $TiS_2$ - as power sources, $MoS_2$, $MoSe_2$ and $SnS_2$ - as catalysts for electrochemical water decomposition [see Ref. 2 and references therein]. Substances containing more sulfur, for example trisulfides, can also have similar layered and reduced dimensional structure. Most transition metals chalcogenides melt incongruently [see Ref. 2 for example]. Therefore, it is difficult to obtain single crystals of these substances by melt techniques such as Bridgman or Czochralski methods. Usually, crystals of these substances are obtained by chemical vapor transport technique and less frequently by the flux technique, whereby a slow cooling of the chalcogenide melt [see Refs. 2-5] is the preferred method. For the vapor transport technique, most often halogens and their compounds are used as transport agents. In this case, there is a possibility of halogen incorporation into the crystal structure of the growing crystal. Hence, for the growth of diselenide crystals free from impurities of other elements, some works use selenium vapors instead as the transport agent. Transport of some dichalcogenides was also studied in [2] without the presence of free chalcogens at high temperatures. ReS$_2$ crystals were grown from the Re$_1$S$_{2.01}$ charge in 900→800°C gradient by Schäfer in several days, as reported in [3]. Therein, it was assumed that sulfur fugacity was minimal while it should be determined by Re/ReS$_2$ equilibrium and be significant at the temperature of synthesis. On the other hand, in other articles Schäfer reported on the possibility of using sulfur vapors as a transport agent [4].

This work described the growth of crystals of several sulfides including TiS$_2$, V$_{1+\delta}$S$_2$, NbS$_2$, TaS$_3$, TaS$_2$, MoS$_2$, WS$_2$, FeS, CoS$_2$, NiS$_2$, PdS and PtS, usually in temperature gradient 800→700°C. Large thin TMDs crystals were grown in the excess of sulfur in 1050→950°C gradient [5]. disulfide and diselenide crystals were obtained by transport in sulfur vapors with the pressure of 9 atmospheres [5, 6]. Previous results have shown that it was possible to obtain disulfide and diselenide crystals but not ditelurides [7-11]. In our previous work [12], we noted that when evaporating selenium or tellurium from the metal and Se (Te) melt at 850-650 °C, only selenium or tellurium evaporated. On the other hand, when sulfur is added to the system, the metal could also evaporate forming mixed dichalcogenide crystals in the cooler part, for example in Nb(Se, S)$_2$. Therefore, the motivation of this work is to study the transport of transition metals and formation of sulfide crystals in sulfur vapors and determine the optimal temperature profile for metal transport in sulfur vapors. The new insights reported in this study allow the design and synthesis of high-quality single crystals of transition metal dichalcogenides, opening the door for high-precision studies of the properties of these systems as well as new types of nanoelectronic applications.

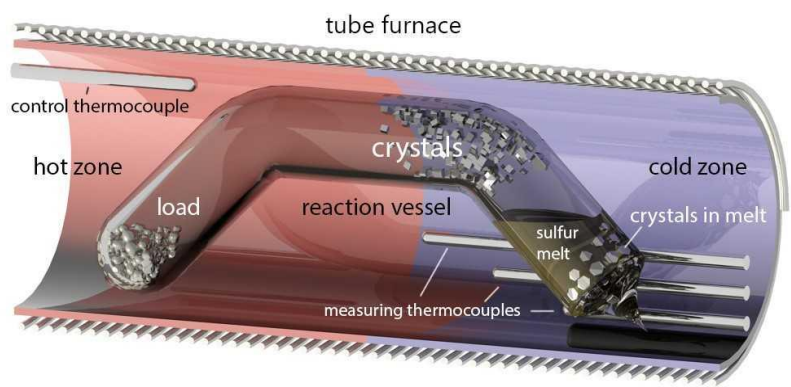

Figure 1. Schematic picture of the reaction vessel (ampoule). Reaction vessel in tube furnace obtaining crystals in quartz ampoules by chemical vapor transport technique with sulfur vapors as the transport agent. The reaction vessel (ampoule) was made from quartz glass tubes 12 mm in diameter with a wall thickness of 2 mm (Fig. 1). The tube had two bends which were made by an oxygen torch. This particular shape of reaction vessel fixed the position of the sulfide charge and the liquid sulfur. About 100-200 mg of metal (or sulfide) and 1-2 g of elementary sulfur were put into the ampoules. Charged ampoules were evacuated and sealed in the flame of the oxygen torch. The total length of the vessels was 150-200 mm. The vessels were placed into the furnace so that the left part containing the sulfide usually had the temperature of 800-850°C and the right part with the liquid sulfur about 550-600 °C. The 550°C temperature of the cooler end provided sulfur fugacity sufficient for substance transport but not enough to destroy the quartz vessel.

**Experimental Section**

Quartz ampoules containing elementary sulfur that are heated to high temperatures are extremely dangerous and unstable. The danger is represented by hot fragments of quartz glass and fumes of sulfur. All manipulations with reaction vessels were carried out with the protection of hands, face and respiratory organs. Sulfur (Labtex 99.9%) and metals with purity no less than 99.9% were used as the reagents. Only transport of those metals whose sulfides did not sublimate at 800-850°C was studied. Therefore, transport of silicon, germanium, cadmium, zinc, tin and mercury was not studied because the crystals of their sulfides are often formed during the synthesis from elements. The 550°C temperature of the cooler end provided sulfur fugacity sufficient for substance transport but not enough to destroy the quartz vessel (see Fig. 1). The highest relative safe temperature of the sulfur part is ~ 640°C at which the pressure reaches 10 atmospheres [13]. This reaction vessel resembles the ones for the three-zone technique described in Schäfer's works. The time scale of the crystal growth ranging between one to four months.

All the experiments conducted to obtain crystals by vapor transport technique with gaseous sulfur are summarized in Table 1 (see below). The table shows the temperature range of coexistence with sulfur, the temperatures of both the hot and the cold ends of the reaction vessel, the approximate growth temperature, the time of the synthesis, the size of the crystals and the transported amount of substance. In case of iron, nickel, niobium and tantalum, complete or almost complete transport of the substance – up to hundred milligrams of metal in 2-4 months through the cross-section of 50 mm$^2$ was observed. In this way $NbS_2$ (Fig. 2a) [14], $FeS_2$ (Fig. 2b), $CoS_2$ (Fig. 2c), $NiS_2$ (Fig. 2d), $Cr_2S_3$ (Fig. 2e), $TiS_2$ (Fig. 2f) crystals of the size up to 2 mm, agglomerates of small crystals of $V_{1+x}S_2$ (Fig. 2g) and $In_2S_3$ (Fig. 2h) and transparent plates MgS (Fig. 3a) were obtained. During the transport of tantalum in close proximity to sulfur source, one-dimensional $TaS_3$ crystals were found up to 30 mm in length and about 1mm thick (Fig. 3b). $TaS_2$ crystals were located in a slightly more high-temperature part (Fig. 3c). The temperature of initial crystallization of $TaS_2$ can be estimated as 675 (50) °C, and that of $TaS_3$ as 625 (50) °C. Similarly, red-orange $ZrS_3$ (Fig. 3d) and $HfS_3$ crystals (Fig. 3e) were obtained in a ribbon shape 10 mm long and up to 0.5 mm wide. In the experiment with platinum 50 mg of $PtS_2$ were obtained in 2 months. The crystals were isometric, a few tens of micrometers in size with well-defined layered structure (Fig. 3f). Whiskers in similar conditions of PdS were formed (Fig. 3g). Growing $PdS_2$ crystals in a similar temperature is likely impeded due to its low temperature of stability.

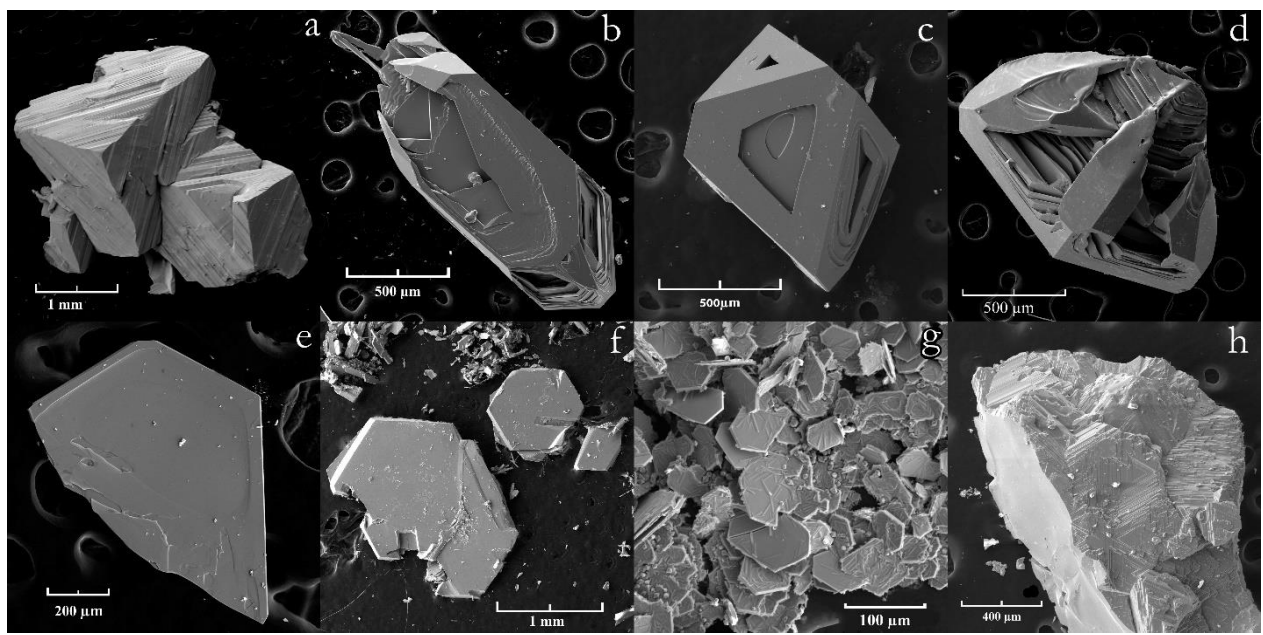

Figure 2. Electron microscope image of crystals: $NbS_2$ (a), $FeS_2$ (b), $CoS_2$ (c), $NiS_2$ (d), $Cr_2S_3$ (e), $TiS_2$ (f), $V_{1+x}S_2$ (g), $In_2S_3$ (h)

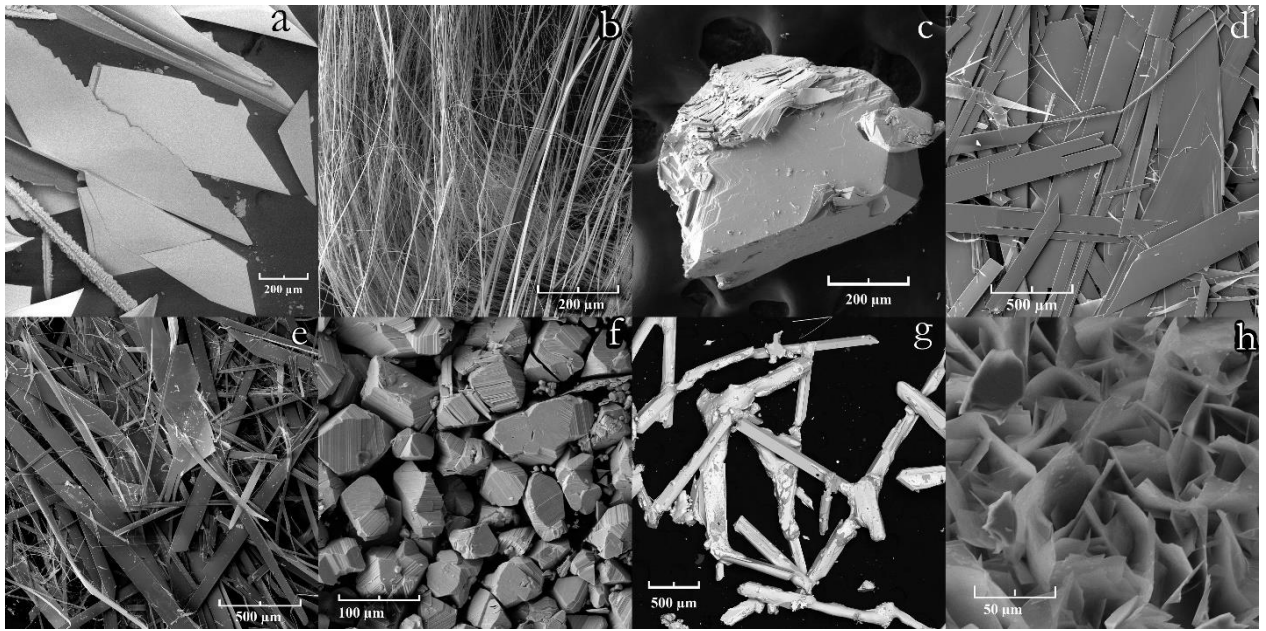

Figure 3. Electron microscope image of crystals: MgS (a), TaS$_3$ (b), TaS$_2$ (c), ZrS$_3$ (d), HfS$_3$ (e), PtS$_2$ (f), PdS (g), MoS$_2$ (h)

Transport of rhenium, molybdenum and tungsten was also observed in similar temperature conditions but significantly less in volume. MoS$_2$ (Fig. 3h) and WS$_2$ crystals a few tens of micrometers in size and ReS$_2$ crystals without a well-defined habit (Fig. 4a) were obtained. Tungsten disulfide crystals located right next to the sulfur source were silver in color. Black and silver areas had well defined borders, therefore it can be assumed that two crystal modifications of WS$_2$ were obtained. The possibility of transferring gold is shown in the standard temperature regime during the recrystallization of the ZnS + Au charge (Fig. 4b). In case of ruthenium, rhodium and osmium no transport was observed. Bi$_2$S$_3$ wiskers (Fig. 4c) were obtained under the most low-temperature regime: charge temperature – 620°C, sulfur source temperature – 540°C. In the experiments to transport palladium, nickel and titanium some additional crystals were found in the sulfur melt, that is, in the coolest part of the reaction system. Most likely gaseous metal compounds dissolved in the liquid sulfur, diffused to the coolest part of the system and formed the crystals there. This way NiS$_2$, Cr$_2$S$_3$ and MgS crystals were obtained with the procedure similar to that obtained by vapor transport. In contrast, during the transport of palladium and titanium, PdS$_2$ (Fig. 4d) and TiS$_3$ (Fig. 4e) crystals were obtained instead of PdS and TiS$_2$, respectively. This can be explained not by different chemical properties of liquid and gaseous sulfur, but by different temperatures of stability of the phases, see Table 1. In the case of transport of manganese, MnS crystals (Fig. 4f) and crystal agglomerates 40 μm in size (Fig. 4g) were found only in liquid sulfur. In order to show that during the transport, the metal enters the gas phase in the form of molecules containing greater amount of sulfur than the charge and the growing crystal (Eq. 1) experiments were conducted without elementary sulfur. The ampoule containing the mixture of NbS$_2$ and a small amount of NbS$_3$ was maintained in 850 → 700°C gradient for two months. The absence of transport in this experiment demonstrates that sulfur fumes are necessary for the transport, which makes this technique different from the sublimation technique.

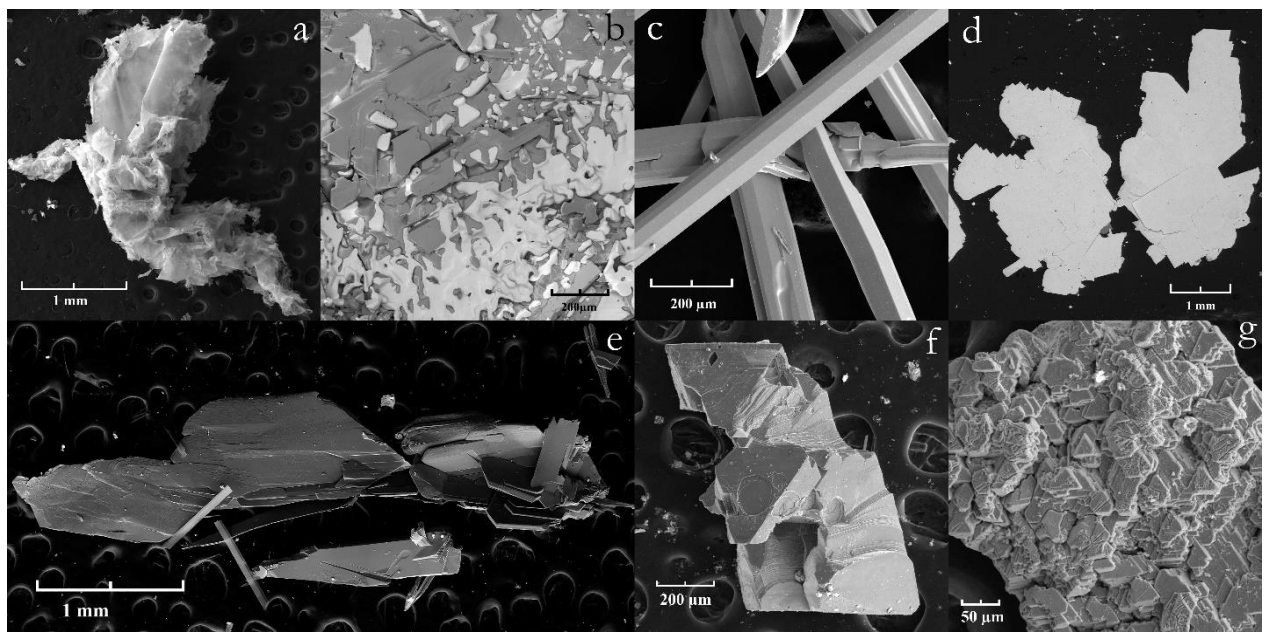

Figure 4. Electron microscope image of crystals: ReS$_2$ (a), Au (white) and ZnS (b), Bi$_2$S$_3$ (c), PdS$_2$ (d), TiS$_3$ €, MnS (f), MnS agglomerates (g)

Chemical composition of the crystals obtained was measured using Tescan Vega II XMU scanning electron microscope with INCA Energy 450 energy dispersive spectrometer in accelerating voltage of 20 kilovolts. Crystals glued to the conducting substrate as well as embedded into polished epoxy resin were studied. The crystals were ground and examined by x-ray powder diffraction technique on DRON-**7** (CoK$_\alpha$-radiation, Fe-filter) or BRUKER (CuK$_{\alpha 1}$-radiation, graphite monochromator) diffractometers. Crystals with apparent layered structure were checked for monocrystallinity on the BRUKER diffractometer. Traces of transport were found almost in all ampoules after 2-4 month in the furnace when the temperature in the hot end was 850-820°C and 570-540°C in the cooler end and if sulfur excess was sufficient. Most often crystals were found closer to the cooler end of the ampoule, that is, the temperature of crystallization was not greater than 650°C.

Hard X-ray photoelectron spectroscopy (HAXPES) was carried out at the KMC-1 beamline using the HIKE end-station [15] located at the BESSY II electron storage ring operated by the Helmholtz-Zentrum Berlin für Materialien und Energie. Photon energies across the S K edge (2465-2490 eV) were selected with a Si (111) double-crystal monochromator and recorded with a Scienta R400 analyzers. The photon beam was at grazing incidence to the sample surface and at normal emission with respect to the analyzer. Magnetization measurements were performed by using a Quantum Design SC quantum interference magnetometer. The low-T specific heat down to 0.4K was measured for TaS$_2$ in its Physical Property Measurement System with the adiabatic thermal relaxation technique. Specific heat measurements were performed down to 70mK by using a heat-pulse technique within a dilution refrigerator along $H \mid\mid c$. We have performed spin-polarized plane-wave pseudopotential calculations with norm-conserved projector-augmented wave (PAW) pseudopotentials, as implemented in the Vienna Ab-initio Simulation Package (VASP) [16]. The exchange-correlation interactions are treated under generalized gradient approximation (GGA) with van der Waals corrections as according to the Grimme DFT-D2 method [17]. The cut-off energy for the plane-wave expansion is set as 500 eV.

**Results and discussion**

X-ray linear dichroism (XLD) studies for ReS$_2$ were carried out at the P64 beamline of the PETRA III synchrotron (Hamburg, Germany) [15]. XLD studies were carried out with the photon incident at grazing incidence in the direction of the incoming X-ray beam. X-ray absorption (XAS) with linearly polarized light is sensitive to the charge distribution and reflects the anisotropy of the electronic states probed in the absorption process. In the case of a non-spherical charge distribution, atoms can produce a different absorption for different orientations of the linear polarization with respect to the sample. For ReS$_2$, the X-ray linear dichroism (XLD) experiments along the *in-plane* direction are shown in Fig. 5(a, b). The fully occupied $t_{2g}^3$ and empty $e_g$ orbitals should show a negligible dichroic signal. However, the spectra show a sizable in-plane dichroic signal (Fig. 5(b)) that point to an orbital anisotropy with a preferential occupation along one direction. This observation can be supported from previous studies on the inter and intra-layer charge occupation of rhenium atoms. This orbital polarization is induced from the additional electron in Re along the strong intra-plane Re-Re metallic bond direction [18].

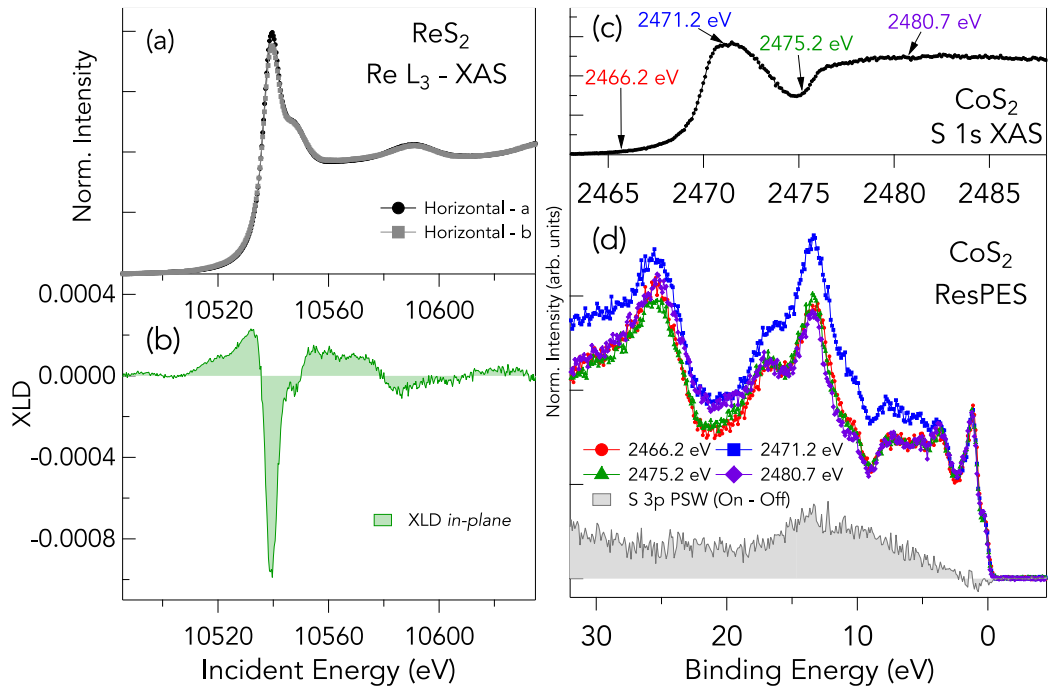

Figure 5. X-ray linear dichroism (XLD) and HAXPES experiments: (a) Re $L_3$ XAS spectra for ReS$_2$ taken at grazing incidence for the two orientations in the *ab* plane. (b) X-ray linear dichroism signal, measured as the difference between the two XAS signals for ReS$_2$ and shows a non-zero signal that points to an anisotropic orbital occupancy within the *ab* plane. (c) XAS spectrum taken at the S $K$-edge for ReS$_2$ with marked energies used for the resonant photoemission spectra (d) Resonant photoemission spectra for CoS$_2$ taken at energies across the S $K$-edge marked in (c). The photoemission spectra taken at 2471.2 eV corresponds to the resonant spectra, while one taken at 2466.2 eV is the non-resonant, and the difference between the resonant and non-resonant spectra represent the partial spectral weight of Sulfur states in the valence band (*grey fill in (d)*), states that lie well below the Fermi level.

Hard X-ray photoelectron spectroscopy (HAXPES) is sensitive to the element-specific density of states in the valence band and shown for CoS$_2$ in Fig. 5(c, d). To map the different spectral contributions of sulfur states in the valence band, resonant photoemission (ResPES) taken at the S $K$ edge probes the S $3p$ states in the valence band. Figure 5(c) shows the S $K$-edge X-ray absorption with marked energies that are used to obtain resonant photoemission spectra shown in Fig. 5(d). CoS$_2$ is known to have a valence band character that is primarily composed of metallic Co $3d$ states at the valence band maximum [19]. The partial spectral weight (PSW, *gray fill* in Fig. 5(d)) shows the difference between resonance and off-resonance condition in the final state that is proportional the S $3p$ states in the valence band. The spectra show a sizable contribution of sulfur $3p$ states that lie several eV from the Fermi level with negligible contribution at the Fermi level. This data confirms the Co $3d$ contribution to the frontier states of the valence band in CoS$_2$ [19-21].

Table 1. Parameters of experiments on sulfide's growth: temperature range of coexistence with sulfur, temperatures of both hot and cold ends of reaction vessel, approximate growth temperature, time of synthesis, size of crystals and carry amount.

| Crystals, crystal structure | Temperature range of coexistence with sulfur, T°C | T°C of hot end (of evaporation) | T°C of cold end (of condensation) | Growth temperature | Synthesis time, days | Crystal size | Carry amount | Fig. № |
|---|---|---|---|---|---|---|---|---|
| $NbS_2$ | n/d | 850 | 550 | ~700-600 | 120 | 1-2 mm | full | 2a |
| $NbS_3$ | n/d | 850 | 550 | ~600 | 120 | below resolution limits | | |
| $PdS$ | n/d | 827 | 551 | ~700-600 | 70 | 2-3 mm × 100 μm | full | 3g |
| $PdS_2$ | n/d | | | 551 (in liquid S) | | 2-3 mm | | 4d |
| $PtS_2$ | n/d | 850 | 550 | ~700-600 | 60 | 50-100 μm | low | 3f |
| $TiS_2$ | > 632 | 820 | 566 | ~700-600 | 93 | 1 mm | full | 2f |
| $TiS_3$ | < 632 | | | 566 (in liquid S) | | 1 mm | | 4e |
| $Bi_2S_3$ | 113 - 775 | 620 | 540 | ~700-600 | 60 | 3 mm × 100 μm | medium | 4c |
| $ReS_2$ | n/d | 820 | 566 | ~600 | 93 | shapeless agglomerates | low | 4a |
| $TaS_2$ | n/d | 820 | 566 | n/d | 93 | 1 mm | full | 3c |
| $TaS_3$ | n/d | | | | | 2-3 mm × 1 μm | | 3b |
| $ZrS_3$ | < 700 | 827 | 551 | ~700-600 | 70 | 10 mm × 0.5 mm | medium | 3d |
| $HfS_3$ | n/d | 827 | 551 | ~700-600 | 70 | 10 mm × 0.5 mm | medium | 3e |
| $MoS_2$ | 115 - 1750 | 820 | 566 | ~700-600 | 93 | 50 μm × 1 μm | low | 3h |
| $WS_2$ (black) | < 400 | 820 | 566 | n/d | 93 | 50 μm × 1 μm | low | |
| $WS_2$ (silver) | > 400 | | | | | 50 μm × 1 μm | | |
| $FeS_2$ (py str.) | 450 – 743 | 827 | 551 | ~700-600 | 70 | 2 mm | medium | 2b |
| $CoS_2$ | 115 - 950 | 827 | 551 | ~700-600 | 70 | 1 mm | low | 2c |
| NiS$_2$ | 115 - 998 | 827 | 551 | + 551 (in liquid S) | 70 | 2 mm | medium | 2d |
| MnS | < 1653 | 800 | 572 | 572 (in liquid S) | 120 | 1 mm + agglomerates of 40 µm crystals | full | 4f 4g |
| V$_{1+x}$S$_2$ | n/d | 800 | 572 | ~700-600 | 120 | agglomerates of 100 µm crystals | full | 2g |
| Cr$_2$S$_3$ | < 1565 | 815 | 560 | ~700-600 | 115 | 1 mm | full | 2e |
| In$_2$S$_3$ | 1090 | 815 | 560 | ~700-600 | 115 | agglomerates of 100 µm | full | 2h |
| *MgS* | n/d | 815 | 560 | ~600 | 115 | 1 mm plates | low | 3a |
| Au +ZnS | 1718 | 815 | 560 | ~700-600 | 115 | agglomerates of 100 µm | medium | 4b |
| RuS$_2$ | n/d | 800 | 572 | - | 120 | - | absent | |
| Rh$_2$S$_3$ or RhS$_3$ | n/d | 800 | 572 | - | 120 | - | absent | |
| OsS$_2$ | n/d | 800 | 572 | - | 120 | - | absent | |

The synthetic routs reported here produce TMDC that can form chare density waves, be superconducting or magnetic, gapped or metallic, and overall with a range of physical properties that reflect the underlying electronic structure [22-24] (Figs. SI 3 and SI 4). In order to make a complete report of the properties of the here synthesized compounds, we discuss in the SI the calculated electronic structure of several transition-metal dichalcogenides (i.e. CoS$_2$, ReS$_2$, NbS$_2$, and TaS$_2$). These systems were selected to provide insight about the interdependence of the crystal structure, chemical composition and the electronic structure (Figs. SI 1 and SI 2). The analysis of the correlation of structural and optical properties of the bulk sulfide single crystals *viz*. CoS$_2$, ReS$_2$, TaS$_2$ and NbS$_2$ can be accomplished by exploring their electronic structures with the help of the first-principles-based density-functional investigation. The Monkhorst-pack grid used for the Brillouin zone sampling of the cubic system is 5×5×5 and for all non-cubic systems, we have used the sampling of 5×5×3. The ionic positions and the lattice parameters are relaxed within the framework of conjugate gradient algorithm until the Hellmann-Feynman force on each ion is less than 0.01 eV/ Å.

**Conclusion**

In summary, we have demonstrated that sulfur vapors with fugacity of approximately 1 atmosphere allow to transport many transition metals and to obtain crystals of sulfides with the maximum possible sulfur content for these conditions. In the absence of sulfur, the transport of these elements was impeded. Normally, crystals grew closer to the cooler part of the ampoule next to liquid sulfur. In several experiments transported metals partly dissolved in liquid sulfur and crystallized as sulfides right therein. Our synthetic routes reported here open for the design and growth of high-quality single crystals for TMDCs research and applications and will allow an unprecedented characterization of the physical properties which leads to better understanding of the underlying mechanisms of TMDCs research.

**Acknowledgment**
M. A. H. acknowledge the financial support from the Swedish Research Council (VR) under project No. 2018-05393. Support by the P220 program of Government of Russia through the project 075-15-2021-604 is acknowledged. D.P. and F.O.L.J acknowledge the financial support from the Swedish Research Council (VR) under project No. 2020-00681 and No. 2020-06409 respectively. T. S. acknowledges financial support from the Swedish Research Council (VR Grant Nos. 2017-05030 and 2021-03675). O.E. acknowledge financial supported by the Knut and Alice Wallenberg Foundation through Grant No.2018.0060. O.E. also acknowledge support by the Swedish Research Council (VR), the Foundation for Strategic Research (SSF), the Swedish Energy Agency (Energimyndigheten), the European Research Council (854843-FASTCORR), eSSENCE and STandUP. The computations/data handling were enabled by resources provided by the Swedish National Infrastructure for Computing (SNIC. Financial support from Vetenskapsrådet (grant numbers VR 2015-04608, VR 2016-05980 and VR 2019-05304), and the Knut and Alice Wallenberg foundation (grant number 2018.0060) is acknowledged. Computations were enabled by resources provided by the Swedish National Infrastructure for Computing (SNIC) at PDC and NSC, partially funded by the Swedish Research Council through grant agreement no. 2018-05973. We thank the Helmholtz-Zentrum Berlin für Materialien und Energie for the allocation of synchrotron radiation beamtime and acknowledge the support by Roberto Felix Duarte during the beamtime.

**References**

[1] Kumar, N. A., Dar, M. A., Gul, R., & Baek, J. B. Graphene and molybdenum disulfide hybrids: synthesis and applications. *Materials Today* **2015**, 18(5), 286-298.

[2] Lyakishev, N.P. (ed.) Phase Diagrams of Binary Metallic Systems: Mashinostroenie, 1996–2000 (in *Russian*).

[3] Wehmeier, F. H., Keve, E. T., & Abrahams, S. C. Preparation, structure, and properties of some chromium selenides. Crystal growth with selenium vapor as a novel transport agent. *Inorganic Chemistry* **1970**, 9(9), 2125-2131.

[4] Späh, R., Elrod, U., Lux-Steiner, M., Bucher, E., & Wagner, S. PN junctions in tungsten diselenide. *Applied Physics Letters* **1983**, 43(1), 79-81.

[5] Dalrymple, B. J., Mroczkowski, S., & Prober, D. E. Vapor transport crystal growth of the transition metal dichalcogenide compounds $Nb_{1-x}Ta_xSe_2$. *Journal of crystal growth* **1986**, 74(3), 575-580.

[6] Legma, J. B., Vacquier, G., & Casalot, A. Chemical vapour transport of molybdenum and tungsten diselenides by various transport agents. *Journal of crystal growth* **1993**, 130(1-2), 253-258.

[7] Al-Hilli, A. A., & Evans, B. L. The preparation and properties of transition metal dichalcogenide single crystals. *Journal of Crystal Growth*, **1972**, 15(2), 93-101.

[8] Schäfer, H. Der Chemische Transport von Re, $ReO_2$ $ReO_3$ und $ReS_2$. *Zeitschrift für anorganische und allgemeine Chemie*, **1973**, 400(3), 253-284.

[9] Schäfer, H., Wehmeier, F., & Trenkel, M. Chemischer Transport mit Schwefel als Transportmittel. *Journal of the Less Common Metals* **1968**, 16(3), 290-291.

[10] Naito, M., & Tanaka, S. Electrical transport properties in $2H-NbS_2$, $NbSe_2$, $TaS_2$ and-$TaSe_2$. *Journal of the Physical Society of Japan* **1982**, 51(1), 219-227.

[11] Barry, J. J., Hughes, H. P., Klipstein, P. C., & Friend, R. H. Stoichiometry effects in angle-resolved photoemission and transport studies of $Ti_{1+x}S_2$. *Journal of Physics C: Solid State Physics*, **1983**, 16(2), 393.

[12] Chareev, D. A., Evstigneeva, P., Phuyal, D., Man, G. J., Rensmo, H., Vasiliev, A. N., & Abdel-Hafiez, M. Growth of Transition-Metal Dichalcogenides by Solvent Evaporation Technique. *Crystal Growth & Design*, **2020**, 20(10), 6930-6938.

[13] Mikkelsen, J. C. PTX phase diagram for Ti−S from (60/75) atomic sulfur. *Il Nuovo Cimento B (1971-1996)*, 1977, 38(2), 378-386.

[14] Cho, C. W. et al. Evidence for the Fulde–Ferrell–Larkin–Ovchinnikov state in bulk $NbS_2$. *Nature Communications* **2021**, 12(1), 1-7.

[15] Gorgoi, M. et al. Nuclear Instruments and Methods in Physics Research Section A: Accelerators, Spectrometers, *Detectors and Associated Equipment* **2009**, 601, 48-53.

[16] Kresse, G. & Furthmüller, J. Efficient iterative schemes for ab initio total-energy calculations using a plane-wave basis set. *Physical Review B,* **1996**, 54, 11169.

[17] Grimme, S. Antony, J. Ehrlich, S. & Krieg, H. A consistent and accurate ab initio parametrization of density functional dispersion correction (DFT-D) for the 94 elements H-Pu. *The Journal of chemical physics,* 2010, 132, 154104.

[18] Gadde, R. et al. Two-dimensional $ReS_2$: Solution to the unresolved queries on its structure and inter-layer coupling leading to potential optical applications. *Phys. Rev. Materials* **2021**, 5, 054006.

[19] Teruya, A. et al. Fermi Surface and Magnetic Properties in Ferromagnet $CoS_2$ and Paramagnet $CoSe_2$ with the Pyrite-type Cubic Structure, *J. Phys.: Conf. Ser.* **2017**, 807, 012001.

[20] Wolfgang A. Caliebe, W. A. et al. High-flux XAFS-beamline P64 at PETRA III. *AIP Conference Proceedings* **2019**, 2054, 060031.

[21] Schroeter, N.B.M. et al. Weyl fermions, Fermi arcs, and minority-spin carriers in ferromagnetic CoS2 *Sci. Adv.* **2020**, 6, 50, 21.

[22] Majumdar, A. et al. Interplay of charge density wave and multiband superconductivity in layered quasi-two-dimensional materials: The case of 2H−NbS$_2$ and 2H−NbSe$_2$. *Phys. Rev. Materials* **2020**, 4, 084005.

[23] Y. Kvashnin, Y. et al. Coexistence of Superconductivity and Charge Density Waves in Tantalum Disulfide: Experiment and Theory. *Phys. Rev. Lett.* **2020**, 125, 186401.

[24] Abdel-Hafiez, M., Zhao, XM., Kordyuk, A. et al. Enhancement of superconductivity under pressure and the magnetic phase diagram of tantalum disulfide single crystals. *Sci. Rep.* **2016**, 6, 31824.